# DEVELOPMENT OF A CONCEPTUAL STRUCTURE FOR A DOMAIN-SPECIFIC CORPUS

*Rushdi Shams, Khulna University of Engineering & Technology (KUET), Bangladesh*
*Adel Elsayed, University of Bolton, UK*

**Abstract**. The corpus reported in this paper was developed for the evaluation of a domain-specific Text to Knowledge Mapping (TKM) prototype. The TKM prototype operates on the basis of both a combinatory categorical grammar (CCG) linguistic model and a knowledge model that consists of three layers: ontology, qualitative and quantitative layers. In the course of this evaluation it was necessary to populate these initial models with lexical items and semantic relations. Both elements, the lexicon and semantic relations, are meant to reflect the domain of the prototype; hence both had to be extracted from the corpus. While dealing with the lexicon was straight forward, the identification and extraction of appropriate semantic relations was much more involved. It was necessary, therefore, to manually develop a conceptual structure for the domain which was then used to formulate a domain-specific framework of semantic relations. The conceptual structure was developed using the Cmap tool of IHMC. The framework of semantic relations- that has resulted from this study consisted of 55 relations, out of which 42 have inverse relations.

## 1 Introduction

The lexicon of TKM prototype developed by Ou & Elsayed (2006) has been populated with lexical items extracted from the corpus developed to evaluate its major components. Efficient parsing of the corpus reflects the richness in linguistic model of TKM prototype but it was inept to map text on its ontology and to represent qualitative and quantitative information due to absence of a conceptual structure for the domain. The corpus contains text that conveys predicate and semantic relations among the elementary data units (Marcu et al, 2001). The predicate relations are useful to populate the lexicon but do not contribute to model the ontology. To model the ontology, we need to identify and categorize semantic relations. Semantic Relations, being qualitative and domain-specific, are important for modeling the ontology and can be formulated from a conceptual structure of the domain (Novak, 2004; Decker et al, 2000).

Being instructional, the text in the corpus sometimes conveys ambiguity to a knowledge mapping prototype if its knowledge model differs from human cognition. For example, a resistor is both a circuit component and a diagrammatic representation. To identify whether the role of a resistor is a component in physical connection or a component in diagram, the machine has to conceptualize the domain like human. A machine only identifies the appropriate roles of concepts in the domain if its knowledge model is developed with domain-specific semantic relations. Semantic relations for a large domain can be obtained by developing conceptual structure of the domain with concept maps as concept maps represent both textual and semantic relations graphically (Nathan & Kozminsky, 2004). Developing conceptual structure of the domain-specific corpus and developing a framework for semantic relations thus are challenges to accomplish.

In this paper, we present a procedure to develop conceptual structure for the domain DC electrical circuit by concept mapping a representative corpus (Mcenery et al, 2006). The corpus currently contains linguistic information like Part of Speech (POS) tags and Combinatory Categorical Grammar (CCG) tags (Clark et al, 2004), and stem of each word- which are useful for empirical linguistics (Lakoff, 1990). These functions enable the corpus aiding the linguist whereas the conceptual structure for the domain aids metacognition of the learners and educators (Fletcher-Flinn & Suddendorf, 1996). We also developed a framework for semantic relations from this conceptual structure which is important for both cognitive and functional linguistics (Gries & Stefanowitsch, 2006).

Anyone unfamiliar with a domain conceptualizes the domain in levels. Starting with reading domain-specific text, the person first conceptualizes the domain by relating the concepts in the text. This cognition is based on predicate relations among concepts. The person needs to relate the concepts with semantic relations if he wishes to extract knowledge represented in the text. This process completes when there remains no other concept except the context- the domain itself. Observing this process of human cognition for a specific domain, we developed the conceptual structure for the domain in levels. We manually conceptualized every sentence in the corpus and then represented them with CmapTools of IHMC (Cañas et al, 2004). These concept maps are at the base level (or level 0) and elicit 55 semantic relations and 42 inverse relations in the corpus. Using FACTOTUM Thesaurus (Chen et al, 2002; Micra, 2008) as a reference framework, we developed a framework to support these 97 domain-specific relations. Afterwards, we analyzed the concepts of the base level, grouped them and linked them with higher level relations to produce level 1 concept maps. This reduces the number of concepts and relations comparing to the base level. In a similar fashion, we developed level 2 concept maps as well. The domain DC electrical circuit is the only concept in context level (level 3). We stopped conceptualizing the domain at that point as we found the context of the domain as the only concept. These four levels of concept maps form the conceptual structure for the domain.

Section 2 depicts the procedure to develop the conceptual structure of the domain from the domain-specific corpus. The section also describes the development of the framework for semantic relations. Section 3 shows the four levels of concept maps for the corpus as well as the framework to support the semantic relations. In section 4, we conclude with a summary and indications for future work.

## 2 The Procedure

This section describes the development procedure of conceptual structure for the domain-specific corpus. We started conceptualizing the domain by taking a sample of the corpus. The concept maps of the sample provide thin predicate relations among concepts. We developed a framework to support predicate relations with a number of semantic relations- which will be used to develop the knowledge model of the TKM prototype.

### *2.1 Conceptual Structure for a Sample of the Corpus*

The corpus contains 1,029 sentences collected from 144 web resources. Initially, we took 308 sentences from the corpus as sample which covers 30 percent of the corpus. We conceptualized each sentence from the sample manually. The outcome of the conceptualization led us to develop concepts and relations among them and graphically represented them with CmapTools. In most cases, the concepts are nouns and the relations are verbs.

To illustrate this procedure, for the sentence *One simple DC circuit consists of a voltage source (battery or voltaic cell) connected to a resistor*, we firstly conceptualized the sentence in the following manner:

1. DC circuit has voltage source as its component.
2. Battery and voltaic cell are voltage sources.
3. Battery and voltaic cell have similarity.
4. Voltage source can be connected to resistor.
5. DC circuit has resistor as its component.
6. As they all are satisfying the properties of a circuit, DC circuit is a type of circuit.

With this conceptualization of the sentence, we then graphically represented the concepts and the relations among them. The concept map for the sentence is depicted in Figure 1.

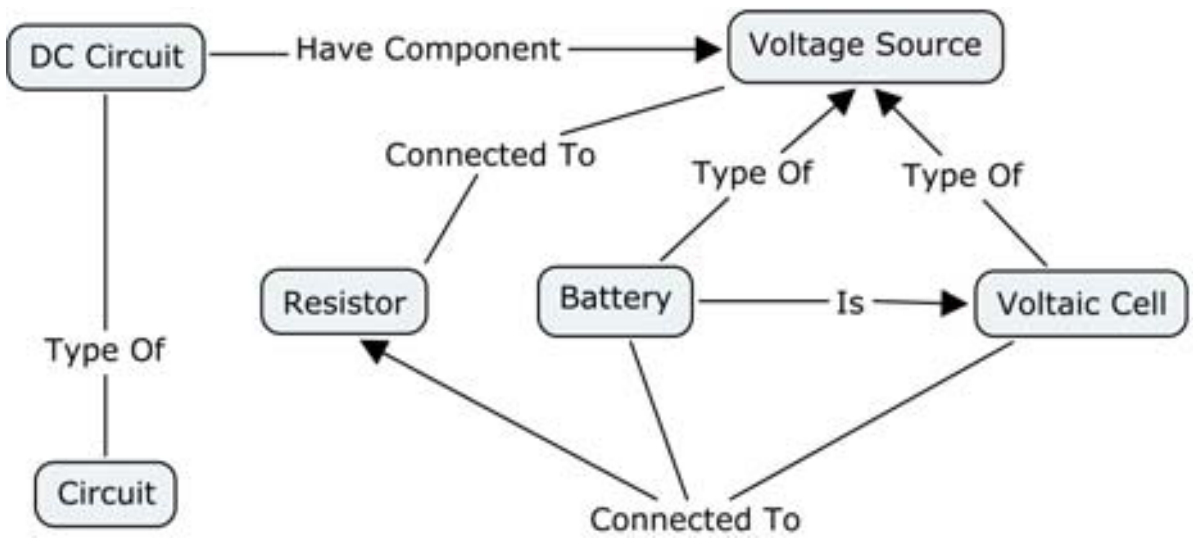

**Figure 1**. Concept map for a sentence from the corpus

As this concept map is developed by conceptualizing a sentence directly, we can say that this concept map is sitting at the base level. To develop higher level concept maps, we require to group concepts according to the semantics embedded in the sentence and to find relationships among these newly created groups. For this particular sentence, we defined groups named *circuit* and *circuit component*. We assigned *DC Circuit* and *Circuit* to the group *Circuit* and the rest of the concepts to the group *circuit component*. We can also find a relation between these two groups- circuit *is made of* circuit components. For a sentence *Resistors in the diagram are in parallel*- the concept *resistor* would be assigned to group of concepts called *Diagrammatic Notation* rather than *Circuit Components*. This process of grouping the concepts from the base level concept maps and finding relations among them produced four levels of concept maps for the sample of the corpus. The conceptual structure of the domain is comprised of all these concept maps resulted from human cognition at four different levels.

*2.2 Framework for Semantic Relations*

The relations that exist in the concept map depicted in Figure 1 are as follows:

1. DC Circuit *Have Component* Voltage source
2. Battery *Type Of* voltage source
3. Voltaic cell *Type Of* voltage source
4. Battery *Is* Voltaic Cell
5. Voltage Source *Connected To* Resistor
6. Battery *Connected To* Resistor
7. Voltaic Cell *Connected To* Resistor
8. DC Circuit *Have Component* Resistor
9. DC Circuit *Type of* Circuit

The relations are completely extracted from the linguistic information carried out by a sentence. They do no help out the user to conceptualize the domain using semantics- which is necessary to extract knowledge from the text. These relations are then analyzed to initiate developing the framework for the semantic relations in the text. The analysis provides us the following relations that are semantically embedded in the text-

1. Predicate Relation which describes parts that are physically related (e.g., Have Component)
2. Predicate Relation which describes hyponymy (e.g., Type Of), and synonymy (e.g., Is) that are similar
3. Predicate Relation which describes hierarchy or class (e.g., Type Of)
4. Predicate Relation which describes spatial relations (specifically location of objects) (e.g., Connected To)

Without creating the concept map from the original text, it is difficult to illustrate the semantic relations embedded in the text. The sentence in concern provides predicate relations that describe parts that are physically related, hyponymy, synonymy, hierarchy, and spatial relation. As the human does acquire and represent knowledge in this way, this process of conceptualization followed by mapping linguistic information on knowledge model also will allow the prototype mapping knowledge from the text onto the ontology efficiently. For example, the prototype now can provide the user knowledge like *voltage source* is a physical part of the *DC circuit*- which is not stated in the sentence but semantically it is present there.

We continue this procedure to reach a stage from where we can constitute a framework for semantic relations. As we continued developing concept maps with the CmapTools, the total number of concepts and relations increases but number of new concepts and relations decreases. We take two pages from the sample as a segment. After every segment, the number of concepts and relations are plotted. Figure 2 shows the cumulative increment of the number of concepts and relations. In sample six, we see a plateau showing that the number of concepts and relations are becoming stationary.

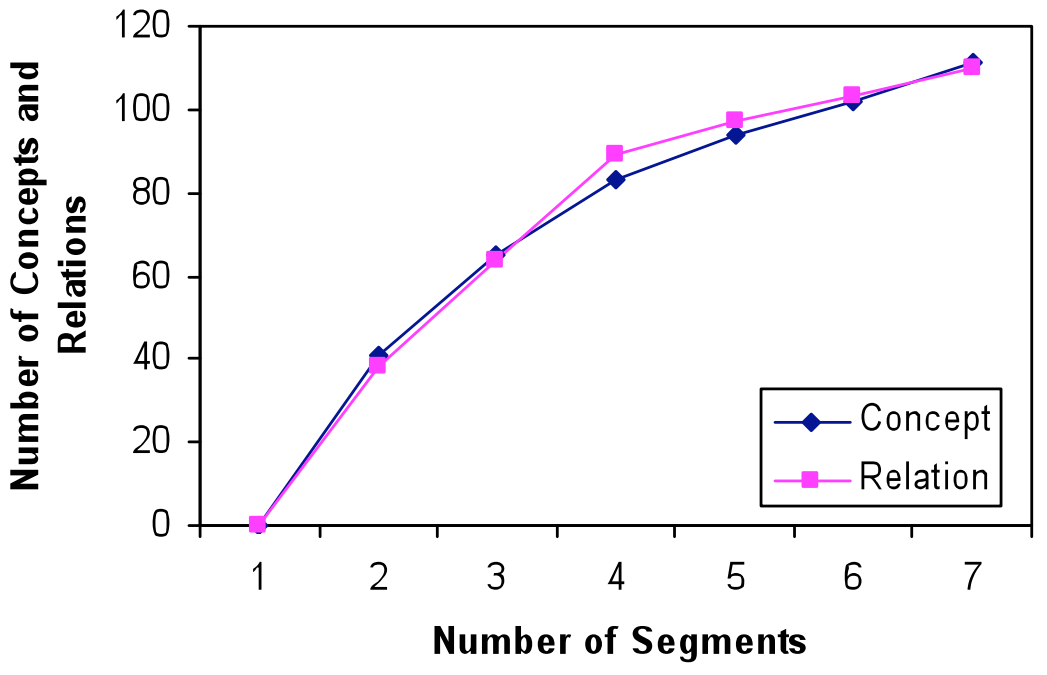

**Figure 2**. Graph to show that the number of concepts and relations in the corpus is becoming stationary

We also plotted number of new concepts and relations found in every segment (Figure 3). The plateau in Figure 3 shows that from sample six, the number of new concepts and relations are becoming stationary. These two observations led us to a decision that at this point (sample six) we can start developing the framework for semantic relations as the number of concepts and relations are not frequently fluctuating. The relations that we will come across by the concept maps after this point can be categorized according to the framework.

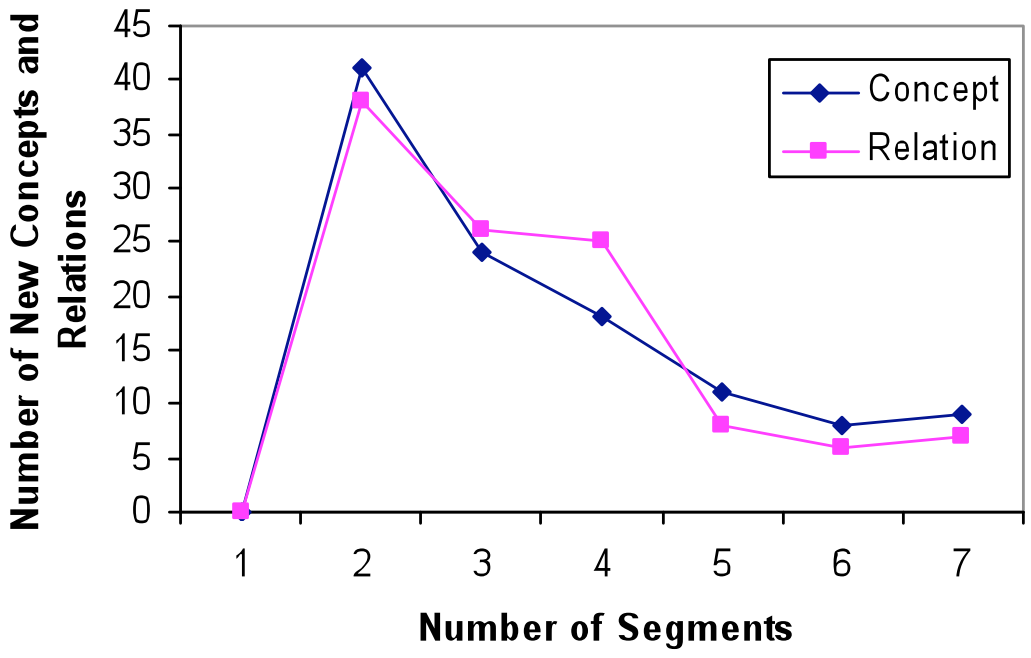

**Figure 3.** Graph to show that the number of new concepts and relations in the corpus is becoming stationary

At this stage, we came across 82 relations and 111 concepts in the sample. Relations in the sample may or may not have an inverse relation. For example, the relation *Have Type* can have inverse relation *Type Of*. In contrast, *Connected To* has no inverse relation. Analyzing all the relations, we found that they are predicate relations and relate the concepts without conveying the semantics. To derive the semantics conveyed by the relations relating the concepts, we developed the semantic relations in Tier 2 (Table 1) and fitted all the 82 predicate relations into the Tier 2 semantic relations.

### *2.3 Conceptual Structure for the Corpus*

After having the relations and concepts from the sample of the corpus, we started developing the concept maps for the whole corpus using them. At stages, we came across new relations and they have been appropriately fitted into Tier 2 of the framework. When base level (level 0) concept maps for the corpus have been developed, there were 97 relations and 166 concepts and we had to adjust Tier 2 to support these relations.

Afterwards, we grouped level 0 concepts and relations to produce level 1 of concept maps. As we came across new predicate relations among concepts, we created Tier 1 to support the semantic relations in Tier 2. These two tiers of semantic relations comprise the domain-specific framework for semantic relations and can be supportive to all the predicate relations of the domain. In essence, the level 0 concept maps have the predicate relations and the semantics conveyed by them are supported by relations in Tier 2. Predicate relations in level 1 and level 2 concept maps are supported by Tier 1 semantic relations.

## 3 Results

Figure 4 shows the concept maps for the whole corpus. There are 12 groups of concepts holding 166 concepts present in the corpus. The concept maps also contain 55 semantic relations and 42 inverses. We call these concept maps- the base level (level 0) concept maps as they are directly developed from the text of the corpus.

From Figure 4, we see that the level 0 concept maps developed from the corpus is not human readable though this level assisted developing Tier 2 of the framework. Therefore, we further grouped level 0 concept maps to develop the level 1 maps shown in Figure 5. For educators and learners, this layer is more appropriate to conceptualize the domain. This level has the same 12 concepts as in Figure 4 but number of relations has been decreased to 11.

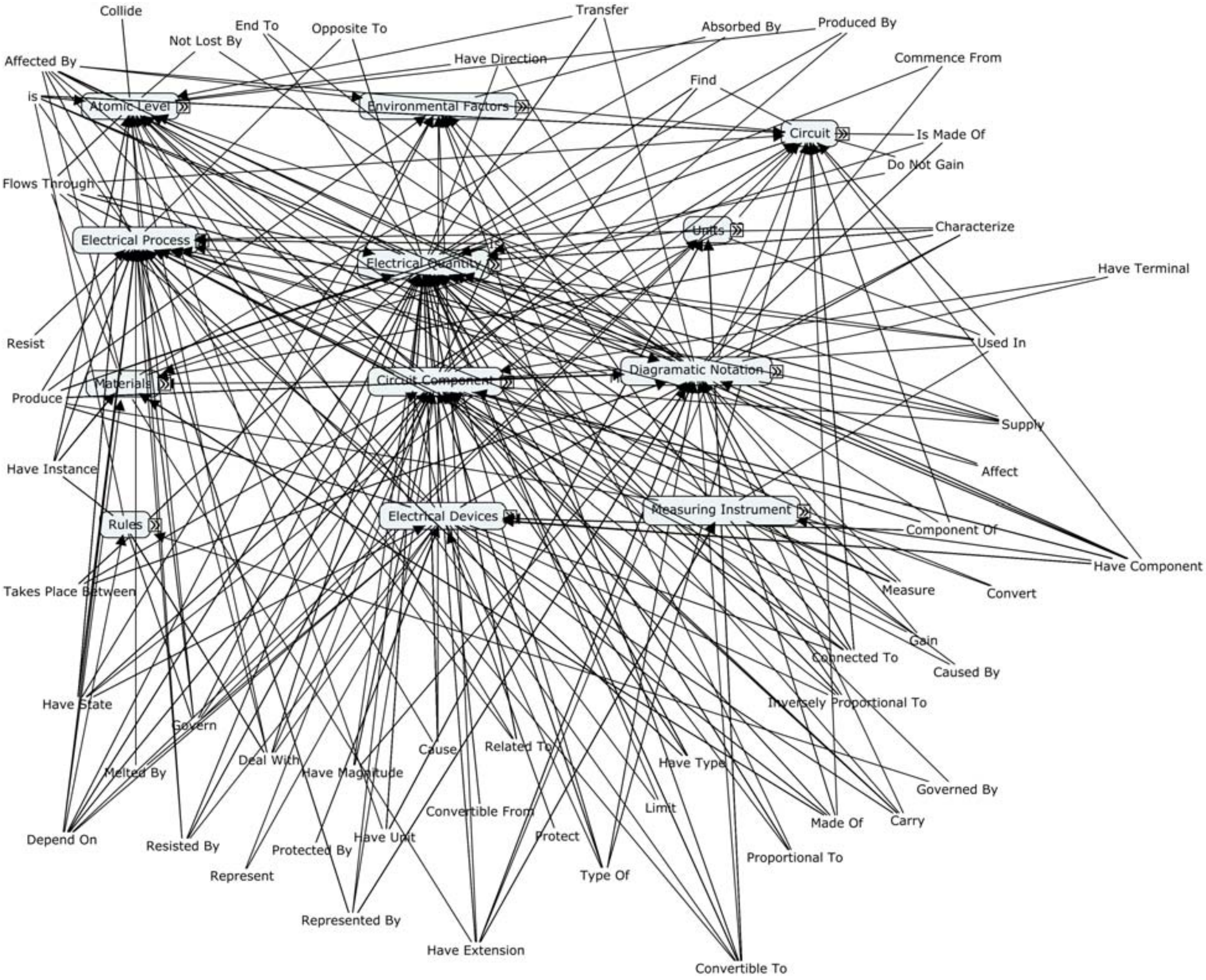

**Figure 4.** Concept maps developed for the corpus

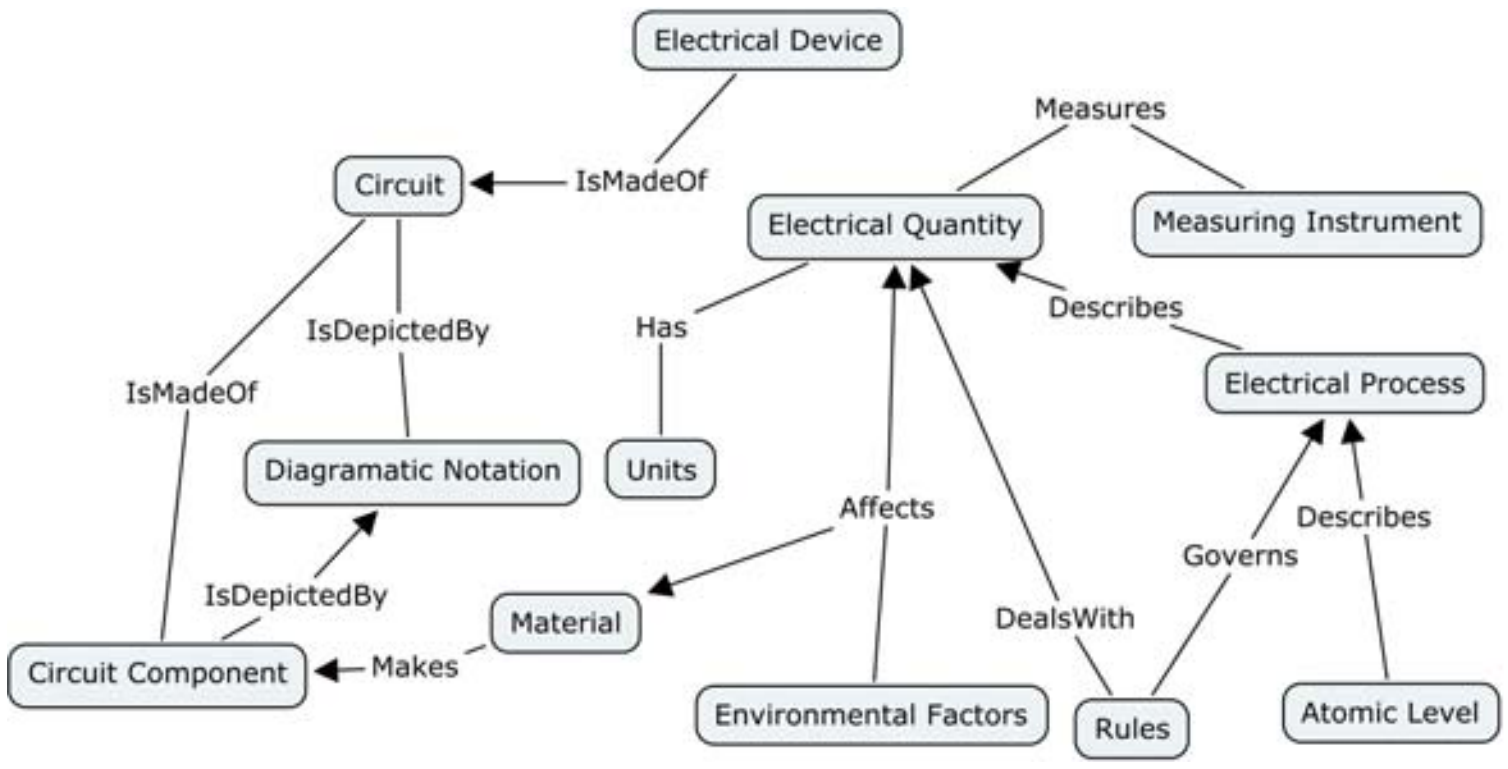

**Figure 5.** Level 1 concept maps for the corpus

Figure 6 shows the level 2 concept maps produced by combining concepts from the level 1 concept maps into groups. In this level, the number of concepts has decreased to six and number of relations has decreased to seven. The concept domain DC electrical circuit alone sits at the contextual level (Figure 7).

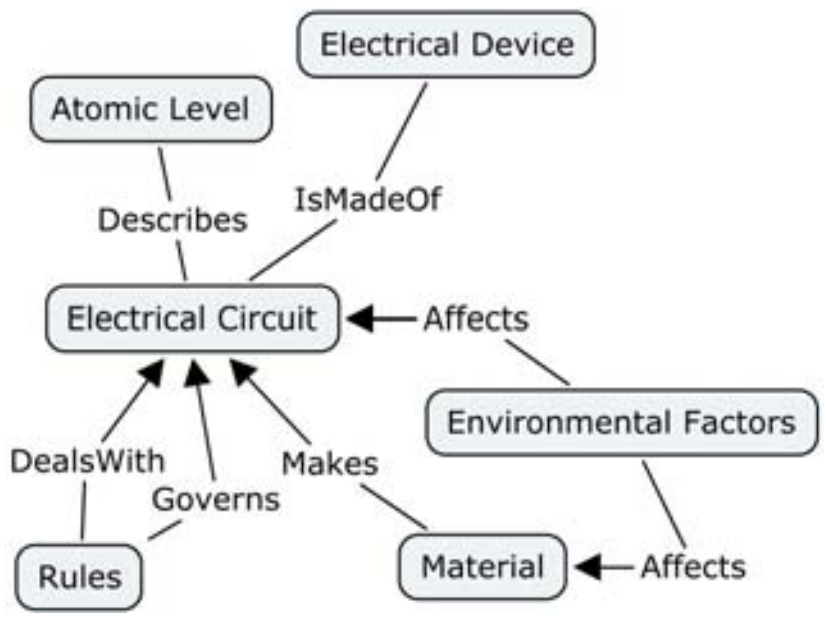

**Figure 6.** Level 2 concept maps for the corpus

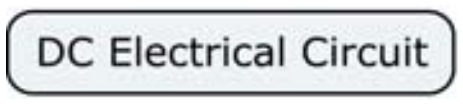

**Figure 7.** Level 3 concept map for the corpus

Therefore, Figure 6 shows the highest level of conceptualization human can have on the domain. Figure 5 shows a more detailed view of the domain and Figure 4 leads the user to the deepest level of conceptualization for the domain. Together these four levels of concept maps form the conceptual structure of the domain.

The domain-specific framework for semantic relations is depicted in Table 1. The framework has three types of relations- predicate relations, instantiation and extension further categorized in three tiers- means the domain-specific corpus has these three semantic relations supported by other relations present in Tier 1 and Tier 2.

| **Semantic Relations** | **Tier 1** | **Tier 2** |
|---|---|---|
| Predicate Relations | Hierarchy/ Class Inclusion | |
| | Physically Related | Parts |
| | | Constituent Material |
| | Spatial Relations | Location of Objects |
| | | Location of Activities |
| | Causally/ Functionally Related | Effect/ Partial Cause |
| | | Production/ Generation |
| | | Destruction |
| | | Manifestation |
| | | Conversion |
| | Instrumental Function/ Usage | Functions |
| | | Use |
| | Human Role | |
| | Conceptually Related | Topic |
| | | Representation |
| | | Property |
| | Similarity | Synonymy |
| | | Hyponymy |
| | Quantitative Relations | Numerical Relations |
| Instantiation | | |
| Extension | | |

**Table 1:** Framework for semantic relations in the corpus

## Discussion and Summary

We developed a conceptual structure for a domain-specific corpus using concept maps. When statistics for the presence of concepts and relations became consistent for a sample of the corpus, we developed Tier 2 of the framework for semantic relations to bridge between predicate and semantic relations among concepts. Then we developed conceptual structure for the whole corpus and discovered new relations among concepts. To support these relations, we developed Tier 1 of the framework. The framework is a generic and a categorical view of the relations existing in large amount of domain-specific text. As the framework is outcome of conceptualization of the domain, it facilitates the prototype for cognitive support and semantic retrieval. Total 97 relations are fit into Tier 2 and 1 that have 16 and nine categories of relations. We developed four levels of concept maps for the corpus that depicts the conceptual structure of the domain. The process of having levels in concept mapping helps cognitive linguistics to pick up the actual semantics embedded in different levels of human cognition. The similar procedure can be applied on any domain to develop a conceptual structure and a domain-specific framework to facilitate knowledge mapping.